\documentclass{pasa}%

\title[Conversion of a New Zealand 30 metre Telecommunications Antenna into a Radio Telescope]{Conversion of a New Zealand 30 metre Telecommunications Antenna into a Radio Telescope}
\author[Woodburn et al.]
{Lewis Woodburn$^1$, 
Tim Natusch$^1$, 
Stuart Weston$^{1,5}$, 
Peter Thomasson$^{1,2}$, 
Mark Godwin$^{3}$, 
Christophe Granet$^{4}$ 
and Sergei Gulyaev $^1$
\\
\affil{$^1$Institute for Radio Astronomy and Space Research, Auckland University of Technology,\\
 Private Bag 92006, Auckland 1142, New Zealand}%
\affil{$^2$Jodrell Bank Observatory, The University of Manchester, Macclesfield, Cheshire, SK11 9DL, UK}%
\affil{$^3$Antenna Measurement and Consultant Services, M P Godwin Ltd., Stoney Middleton, Hope Valley, Derbyshire S32 4TQ, UK}%
\affil{$^4$BAE Systems Australia Ltd., Evans Building, Taranaki Road, Edinburgh Parks, SA 511, Australia}%
\affil{$^5$Email: sweston@aut.ac.nz}}%
\jid{PASA}
\doi{10.1017/pas.\the\year.xxx}
\jyear{\the\year}

\usepackage{graphicx}
\usepackage{epstopdf}
\usepackage{float}
\usepackage{url}
\usepackage{tikz}
\usepackage{subfigure}
\usepackage{wrapfig}
\usepackage{multirow}

\usepackage[authoryear]{natbib}
\bibpunct{(}{)}{;}{a}{}{,}
\begin{document}%
\begin{abstract}
The conversion of a former 100-foot (30-m) telecommunications antenna (Earth Station) in New Zealand into a radio telescope is described. A specification of the antenna and the priorities for its actual conversion are initially presented. In describing the actual conversion, particular emphasis is given to mechanical and electrical components, as well as to the design of the telescope control system, telescope networking for VLBI operations, and telescope maintenance.  Plans for RF, front- and back-end developments based upon radio astronomical priorities are outlined.
\end{abstract}
\begin{keywords}
Antenna -- Conversion -- Radio Telescope -- Instrumentation -- VLBI 
\end{keywords}
\maketitle%
\section{INTRODUCTION }
\label{sec:intro}

\begin{figure}
    \centering
    \includegraphics[width=8cm]{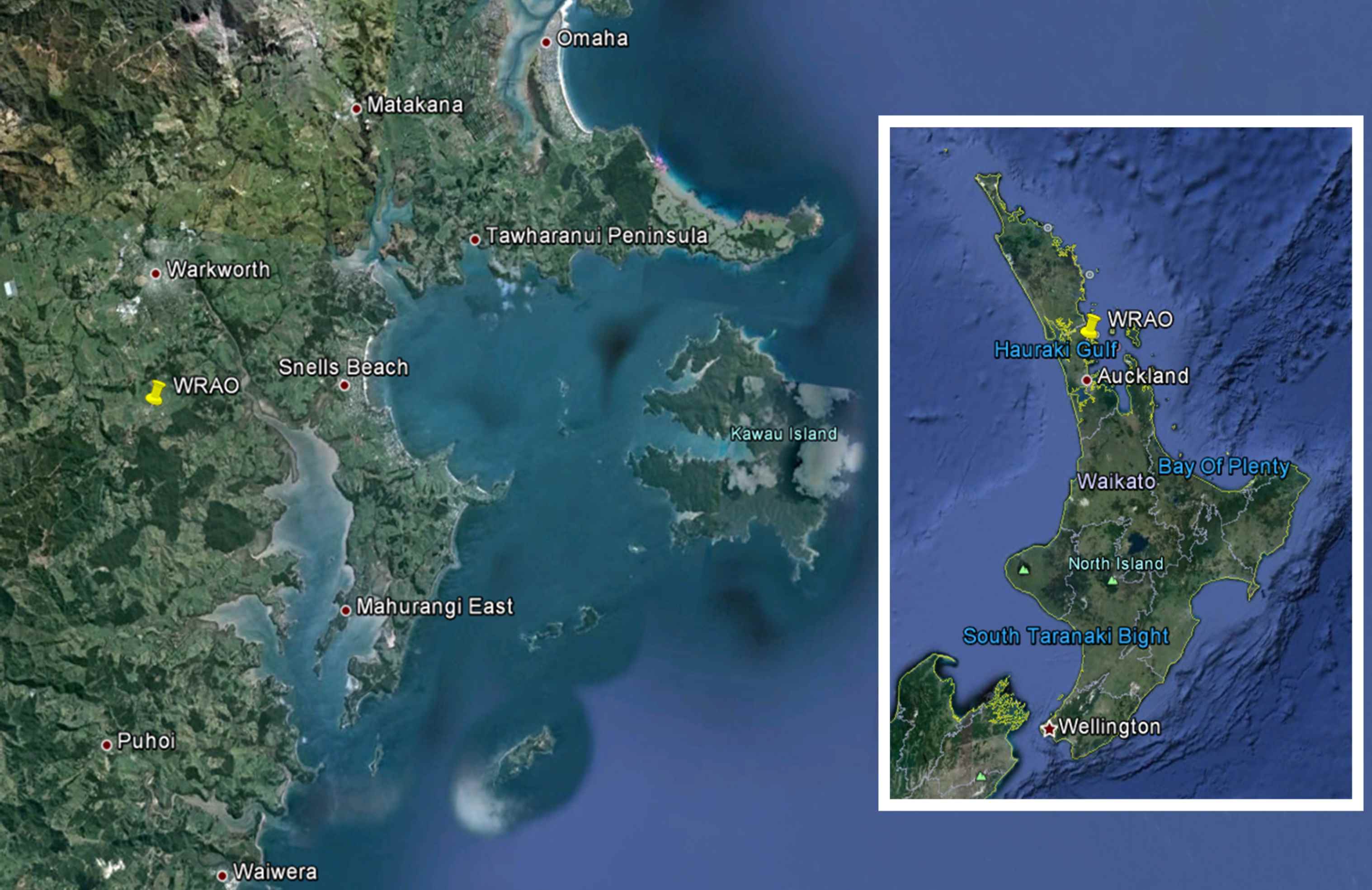}
    \caption{Geographic location of WRAO. The insert shows a map of New Zealand`s North Island with location of WRAO.  (Google Earth)}
    \label{fig:location_image}
\end{figure}

A 100-foot (30-m) Earth Station, located approximately 60~km north of Auckland and 5~km south of Warkworth in the North Island of New Zealand, was designed and built in 1984 by the NEC Corporation, Japan, for the New Zealand Post Office,  who operated it until 1987 when its operation was handed over to Telecom New Zealand, a company formed from the telecommunications division of the New Zealand Post Office. In 2010 it was transferred to the Institute for Radio Astronomy and Space Research (IRASR) of Auckland University of Technology  (AUT) for conversion to a radio telescope \citep{CellularNews:19112010, NBR19112010}. At that time IRASR was already operating a 12-m radio telescope on an adjacent site, which had been officially opened at the Warkworth Radio Astronomical Observatory (WRAO) in 2008. Its location is shown in Figure~\ref{fig:location_image}. 

\begin{figure*}
    \centering
    \includegraphics[width=17cm]{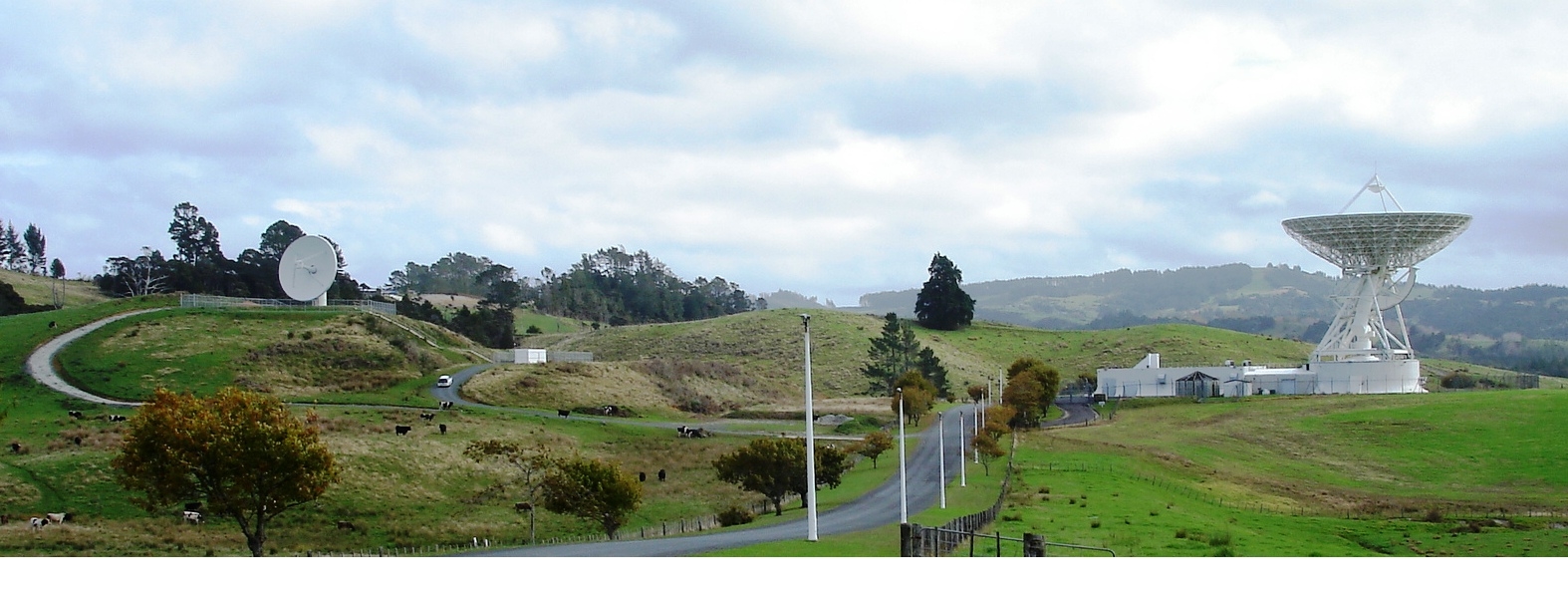}
    \caption{Shows a panorama of the WRAO: 12-m radio telescope is on the left, 30-m is on the right. (Image courtesy of Sergei Gulyaev)}
    \label{fig:wrao_image}
\end{figure*}

Currently available geographic coordinates for the 30\,-m antenna are Longitude: 174$^{\circ}$ 39' 46" E; Latitude: 36$^{\circ}$ 25' 59" S, and Altitude: 90 m.  Figure~\ref{fig:wrao_image} shows a panorama of the WRAO, the horizontal distance between the 12-m (left) and 30-m (right) antennas being $\sim$ 200 m.

The 30-m antenna is not the first satellite-communications antenna to be converted to a radio telescope. The 30-m dish at Ceduna in Australia \citep{McCulloch_2005AJ} was taken over in 1985 by the University of Tasmania for use as a radio telescope, and  a similar conversion was undertaken for a 30-m satellite-communications antenna near Atlanta, United States when it was acquired and fully renovated for radio astronomy use by the Georgia Institute of Technology \citep{Deboer_1999}. A 32-m antenna at Yamaguchi in Japan \citep{Fijusawa_2002}  was taken over by the National Astronomical Observatory of Japan (NAOJ) in 2001 for use by NAOJ and Yamaguchi University and similar projects at Earth Station sites at Goonhilly in the UK \citep{2011arXiv1103.1214H} and Elfordstown near Cork in the Republic of Ireland \citep{2005ASPC..340..566G} are also being considered, which would enhance the resolution and $uv$-coverage of both the Multi-Element Radio-Link Interferometer Network (e-MERLIN) and the European Very Long Baseline Interferometer (VLBI) Network (EVN). With the location of a part of the Square Kilometre Array (SKA) in Africa, there is now considerable interest in radio astronomy there, and recent conversions of former, satellite-communications antennas in Africa \citep{Nordling_2012, physicsworld_10_2012, gaylard:2012} to produce an African VLBI network have been reported. Some of these African antennas, such as at Kuntunse in Ghana, are very similar to the Warkworth 30-m in design and structure.

An initial report on the conversion of the Warkworth 30-m antenna was given in Physics World \citep{physicsworld_12_2012} and a subsequent progress summary was published in the International VLBI Service for Geodesy \& Astrometry (IVS) 2012 annual report \citep{IVS_Annual_Report_2012}. It is now possible to provide a more detailed report on the conversion of this antenna, which is now fully steerable and has been used at C-band ($\sim6.7$~GHz) for commissioning observations using an un-cooled receiver. 

\begin{figure}
\centering
\begin{subfigure}
  \centering
  \includegraphics[width=0.22\textwidth]{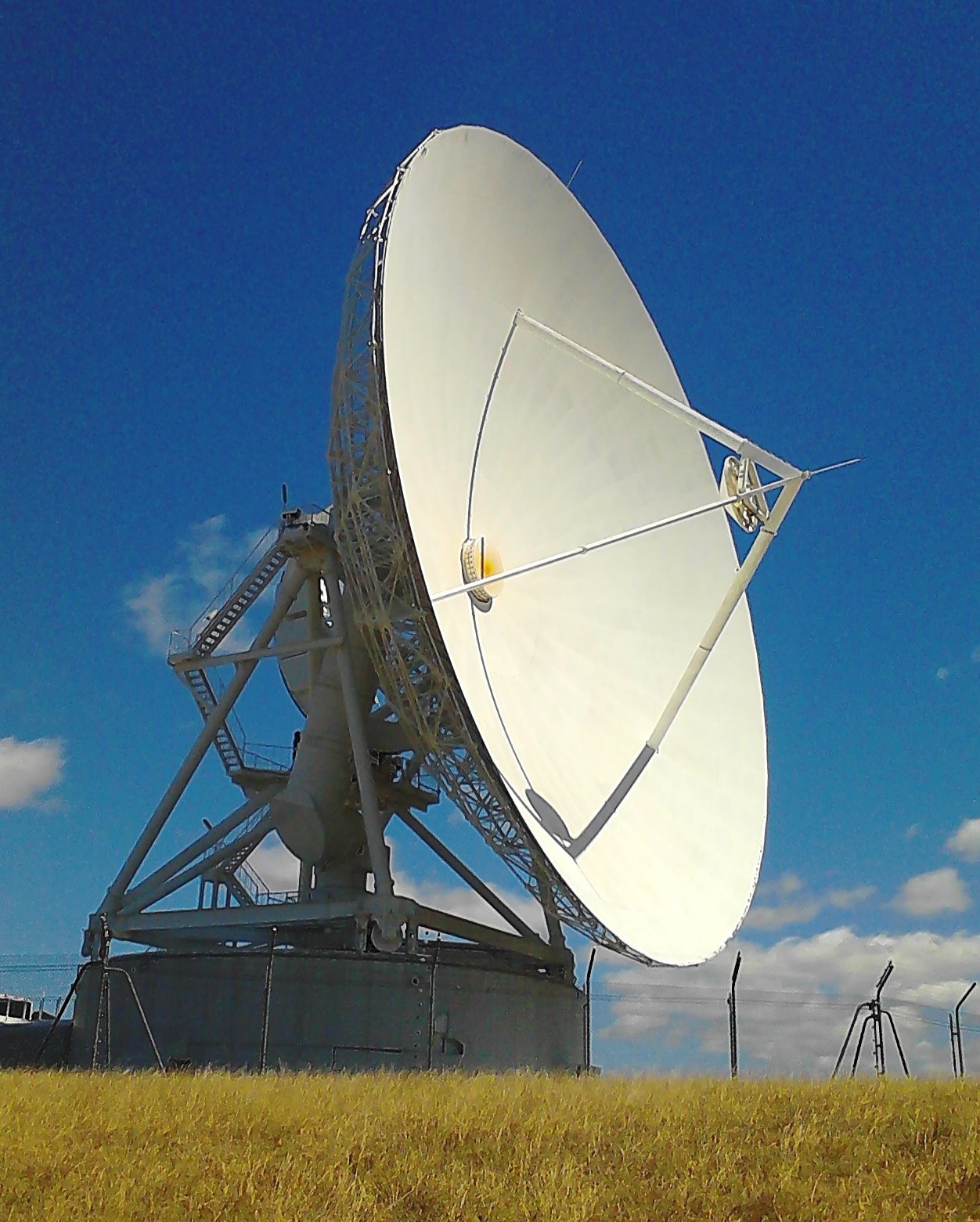}
   \label{fig:sub1}
\end{subfigure}%
\begin{subfigure}
  \centering
  \includegraphics[width=0.21\textwidth]{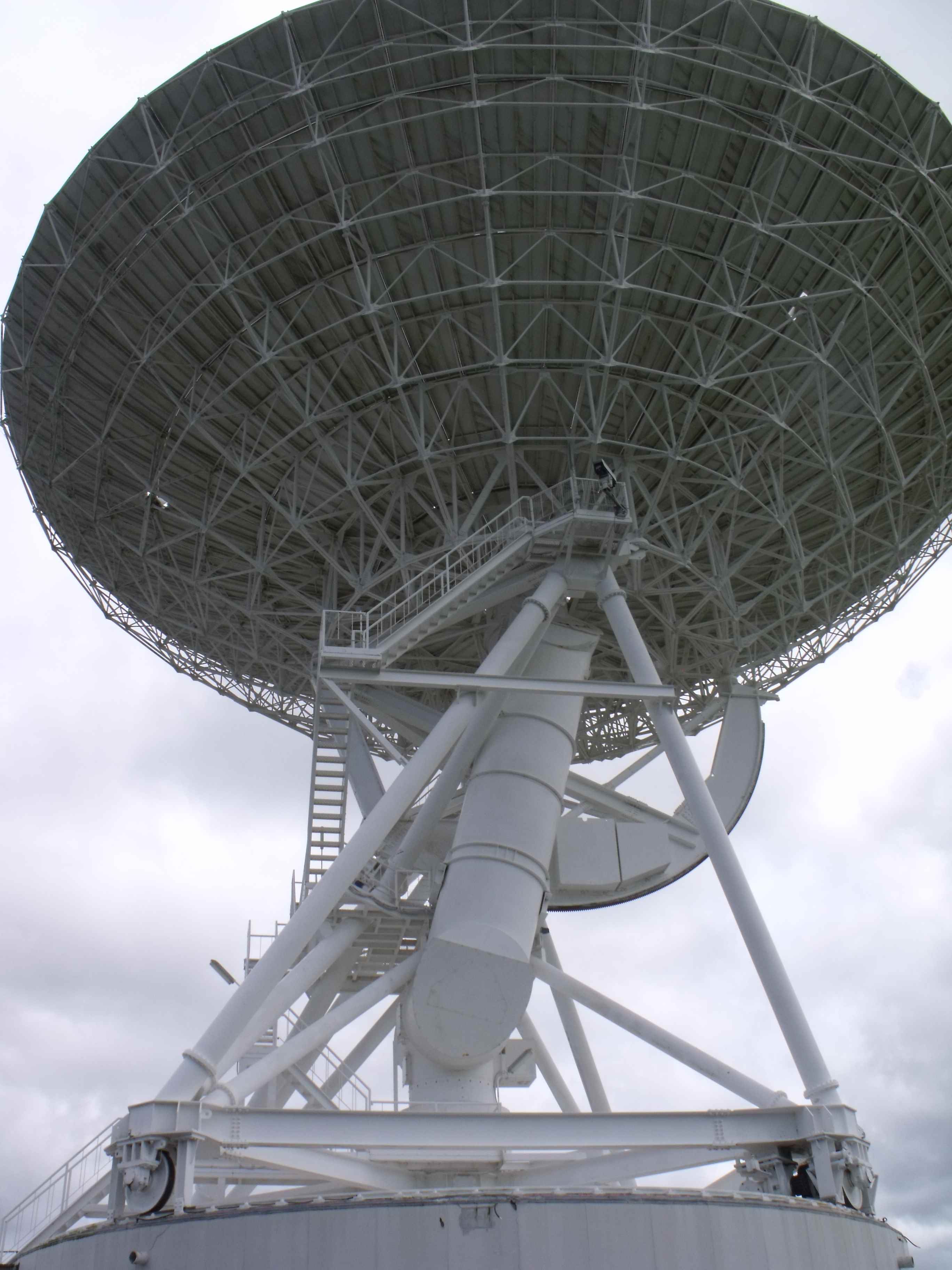}
  \label{fig:sub2}
\end{subfigure}
\caption{Photographs of the 30-m antenna after cleaning. (Images courtesy of Stuart Weston)}
\label{fig:30mClean}
\end{figure}

\begin{table*}
\caption{Specifications of the Earth Station according to the manufacturer`s (NEC) handbook.}
\begin{center}
\begin{tabular*}{\textwidth}{@{}ll@{}}
\hline\hline
Description & Specification \\
\hline%
 System & Alt-azimuth, wheel-and-track, Cassegrain, \\ & beam-waveguide antenna \\
Drive system & Electric-servo, dual train for antibacklash \\
Transmission frequency band & C-band \\
Reception frequency band & C-Band \\
Primary mirror diameter& 30.48 m \\
Subreflector diameter & 2.715 m  \\
Azimuth Maximum Velocity in Slew Mode (Open Loop) &  0.3 deg/sec or 18.0 deg/min \\
Elevation Maximum Velocity in Slew Mode (Open Loop) & 0.3 deg/sec or 18.0 deg/min \\
Max Acceleration/Deceleration in both axes & 0.2 deg/sec/sec \\
Max Tracking Velocity (Closed Loop) & 0.03 deg/sec or 1.8 deg/min \\
 & (estimated as no data in the NEC documentation)\\
Azimuth Working Range (as defined by soft limits) & $-170^\circ$  to $170^\circ$ \\
Elevation Working Range (as defined by soft limits) & $0^\circ$ to $90^\circ$ \\
Surface accuracy (rms) & 0.4 mm \\
Track diameter & 16.97 m \\
Total weight on track & 268 tons \\
Wind speed in tracking operation & up to 40 m/s \\ 
 Survive wind speed in stow position & up to 70 m/s \\
\hline\hline
\end{tabular*}
\end{center}
\label{tab:nec_spec}
\end{table*}

Table \ref{tab:nec_spec} presents the original specifications of the Warkworth 30-m Earth Station according to the manufacturer`s (NEC) handbook. Photographs of the antenna (after cleaning - see section \ref{cleaning}) are shown in Figure~\ref{fig:30mClean}.  

\section{PLAN FOR CONVERSION}

After a thorough inspection of the facility and following several discussions, a plan for the antenna conversion was drawn up, which included the following items:-
\begin{itemize}
   \item Replacement of the motors, cables, drive, control system and encoders in order to bring the telescope up to the required mechanical performance.
   \item Design and installation of a cable twister, and modification to the positions of the limit switches.
   \item Cleaning of the surfaces of the antenna and its supporting structure, treatment of rusty surfaces, replacement of rusty bolts and painting.
   \item Implementation of software to enable the telescope to track sources in a celestial co-ordinate system.
   \item Initial installation of a C-band receiver so that first observations could make use of the existing feed system.
   \item Telescope pointing and sensitivity measurements at C-band. (A requirement for this was the inclusion of a noise-diode switching system and a total-power ``back-end''.)
   \item Installation of time and frequency standards (derived from the 12-m telescope).
   \item Purchase of a new Digital Base Band Converter (DBBC) and VLBI recorder.
\end{itemize}

\begin{figure*}
   \centering
   \includegraphics[width=\textwidth]{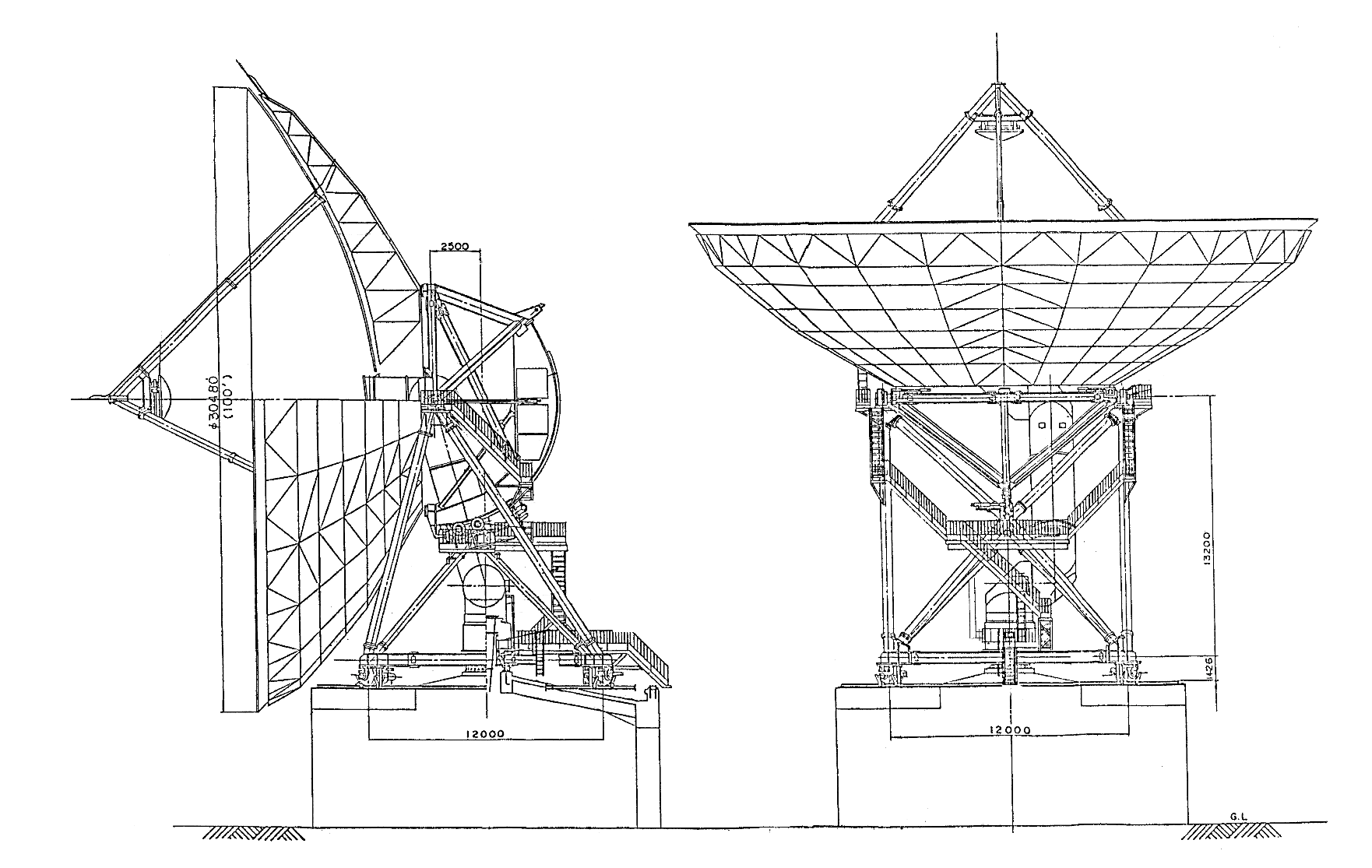}
   \caption{Line drawings of the Warkworth 30-m radio telescope - based on a modified NEC drawing,1984. Measurement units are mm. }
   \label{fig:warkworth2_line_drawing}
\end{figure*}

\section{IMPLEMENTATION OF THE CONVERSION PLAN}

A line drawing of the 30-m antenna is shown in Figure~\ref{fig:warkworth2_line_drawing}. In addition to the plan items listed above, although the NEC surface accuracy specifications  (rms = 0.4 mm) indicated that the telescope should be operational at frequencies up to at least 22 GHz and probably higher, after many years of operation and then inactivity it was unclear that this was still the case. Therefore measurements of the current state of the surface and variations of it with elevation also became an intrinsic part of the plan. These measurements are described in Section \ref{surface}.
 
\subsection{Azimuth drive limits, cables and motors}

For the efficient implementation of most radio astronomy telescope schedules, azimuth movements in excess of $360\,^{\circ}$ are essential. The original Earth Station azimuth motion was limited to $\pm 170\,^{\circ}$ relative to zero (due North) and it was decided that a total azimuth movement of $540\,^{\circ}$ ($\pm 270\,^{\circ}$) would be appropriate, making installation of a new cable wrap or twister necessary (See section \ref{cable-wrap}). This, of course, meant a significant increase (in some cases a very significant increase) in the length of the cables passing between the fixed and moving parts of the structure. It rapidly became clear that all the old cables needed to be replaced and new cables for the new drive and control system were required. 

\begin{figure}[!h]
\centering
\begin{tabular}{c}
\includegraphics[height=7cm]{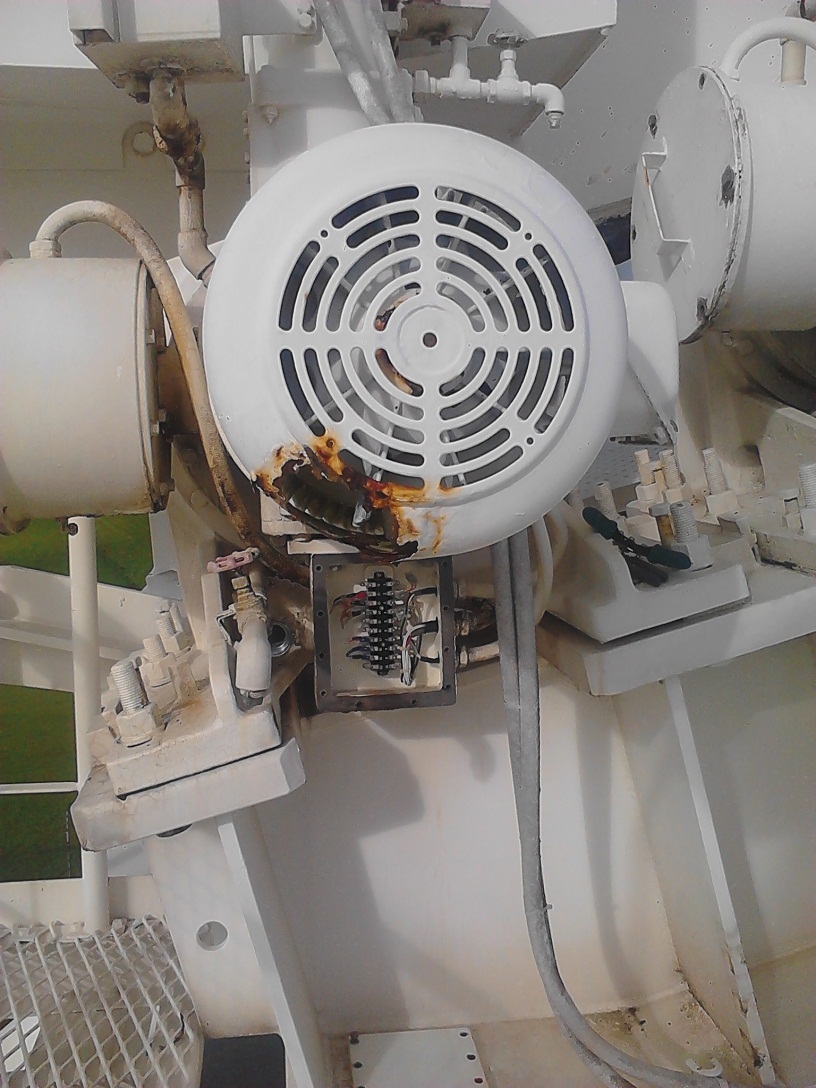}
\end{tabular}
\caption{The state of the old motors was very poor and was considered a safety issue. (Image courtesy of Stuart Weston)}
\label{fig:old_motor}
\end{figure}

It was also decided that the original old NEC motors (one of the original elevation motors is shown in Figure~\ref{fig:old_motor}) should be changed to brushless permanent magnet servomotors to allow the bi-directional 0 to 3000 rpm drive systems required for astronomical operation. Replacement of the motors was also dictated by health and safety issues because of the very rusty
condition of their outside shells. Both azimuth and elevation drives now use two motors in an anti-backlash configuration. Four motors plus one spare of the same model (190U2H200BASAB15420-8RAS) were manufactured by Control Techniques Dynamics (UK) Ltd.  with extended drive shafts of the same diameter as the old NEC motors.

\subsection{Position encoders }

Originally a pair of coarse and fine resolvers was used on each axis to measure angle.  However, resolvers do not provide sufficient accuracy for telescope use and are not readily compatible with modern digital control systems.  Hence they were replaced with absolute optical position encoders.

The new elevation 26 bit encoder, a Heidenhain ROC226, provides a specified absolute accuracy of $\pm 5$ arcsec or $\pm 0.0014^\circ$.  It is mounted in place of the old resolver unit and is driven directly from the elevation shaft via a flexible coupling.  

The beam waveguide configuration does not provide a convenient azimuth shaft from which an encoder can be driven directly.  In the original arrangement a 600 tooth ring gear around the feedhorn drove the azimuth resolver assembly via a 30 tooth anti-backlash pinion.   Thus the input shaft of the dual resolver system rotated 20 times for each revolution in azimuth. The new arrangement retains the precision 600 tooth ring gear, but this now drives the optical azimuth encoder, again via a 30 tooth anti-backlash pinion.  The 25 bit azimuth encoder, a Heidenhain ROQ437, provides a 12 bit turn count and the absolute accuracy of $\pm 20$ arcsec or $\pm 0.006^\circ$.  On the azimuth axis the contribution to the error in the measurement angle is only one twentieth of this value, approximately $\pm 0.0003^\circ$, because of the gear ratio.  However the gears themselves introduce other errors because of minor misalignments and machining tolerances.  No estimate of the magnitude of these errors is available, but pointing calibrations confirm an overall satisfactory system performance.

\subsection{Limit Switches}
To protect the antenna from the damage that would result from driving past known safe limits, a three level hierarchy of limit switching is implemented --- see Table~\ref{tab:switch_limits}.

\subsubsection{Soft Limit Switches}
In the normal course of operation the antenna controller decelerates the telescope motion when approaching certain limits of movement and brings the antenna to a stop at the user set ``soft limits" (Table~\ref{tab:switch_limits}). Should a fault occur that results in the antenna moving past these soft limits, the following hierarchy of limits operate to prevent damage.

\begin{figure}[h]
\centering
  \includegraphics[width=0.47\textwidth]{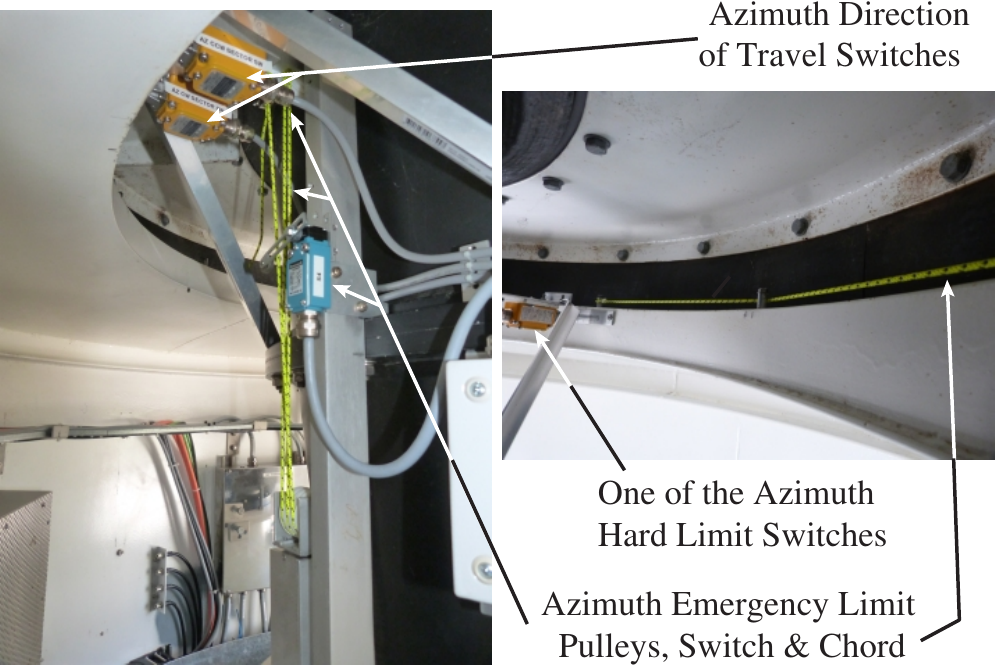}
\caption{Photos of the azimuth emergency limit switch arrangement. The left photo shows the pulley arrangement, and the right photo is the cord playout around the inside of the cable wrap room ceiling. Also shown in the left-hand image are the azimuth direction of travel swiches;  one of the azimuth hard-limit switches is shown in the right-hand image. (Images courtesy of Stuart Weston)}
\label{fig:AzEmLimSwitch}
\end{figure}

\begin{figure*}
    \centering
    \includegraphics[width=15cm]{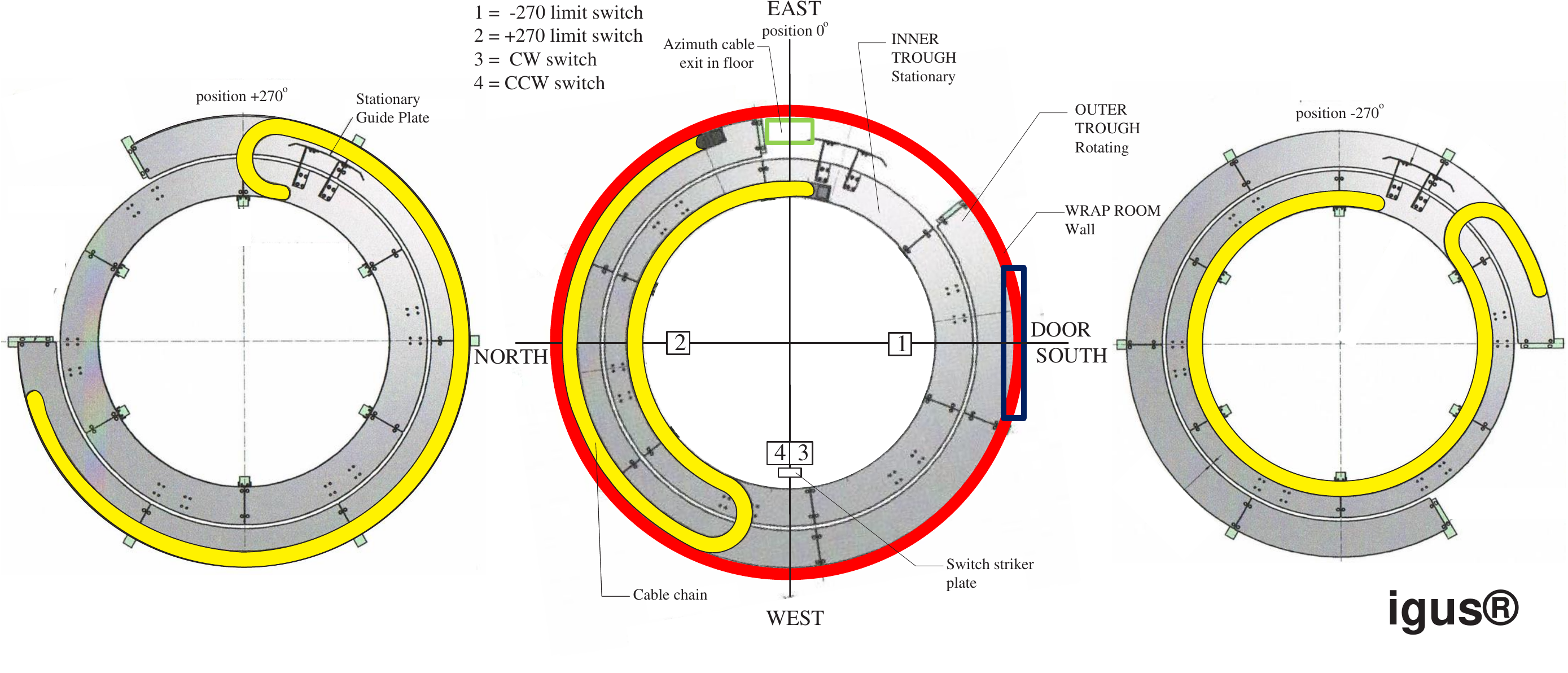}
    \caption{The IGUS Energy chain system drawing as installed. Drawing courtesy of IGUS with added notation by L.Woodburn.}
    \label{fig:EnergyChainsystem}
\end{figure*}     

\subsubsection{Hard Limit Switches}
\underline{Azimuth}.  In the original installation the azimuth hard limit switches were also mounted under the floor of the cable wrap enclosure, which necessitated wiring to be passed through the cable wrap. After servicing, the original switches were re-installed in a position under the cable wrap room ceiling and fixed to the central RF Feed Housing, eliminating the need for the cable to pass through the cable wrap and thus allowing easy access for adjustment, as can be seen in the right hand image of Figure~\ref{fig:AzEmLimSwitch}. 

\underline{Elevation}. The original elevation hard limit switches were serviced and re-installed in the same positions as previously.

\begin{table}
\caption{The limit switch scheme.}
\begin{tabular*}{\columnwidth}{@{}lcccc@{}}
\hline\hline
\multirow{2}{*}{Switch} &   \multicolumn{2}{c}{Low limit} & \multicolumn{2}{c}{High limit} \\
                                     & Az    & El                                   &  Az  & El        \\
\hline%
Soft limit           &  $-179.0^{\circ}$    & $6.0^{\circ}$   & $354.0^{\circ}$   & $90.1^{\circ}$ \\
Hard limit        &    $-179.3^{\circ}$    & $5.7^{\circ}$   & $354.3^{\circ}$   & $90.4^{\circ}$ \\
Emergency stop &  $-180.3^{\circ}$    & $4.7^{\circ}$   & $355.3^{\circ}$   & $91.4^{\circ}$ \\
\hline\hline
\end{tabular*}
\label{tab:switch_limits}
\end{table}

\subsubsection{Emergency limit switches}

On each of the azimuth and elevation axes final mechanical emergency limit switches have been configured to ensure the antenna is brought to a stop in the event that it travels past one of the relevant mechanical hard stop switchs by more than $1^\circ$ (Table~\ref{tab:switch_limits}). These “Emergency Stop” switches are hardwired into the antenna controllers' Emergency Stop circuits. If activated, they open the main power contactor and engage the motor brakes, which are only disengaged if power is present. The telescope control is then switched to the maintenance mode, when the telescope can be driven out of the limit manually.

\underline{Azimuth}.  As the antenna now operates through $\pm 270\,^{\circ}$ in this axis, it is not possible to use switches positioned after the mechanical  limit switches. By using a number of pulleys and nylon sail cord a system has been created that operates a switch that is mounted on the non-moving central RF Feed Housing. The nylon sail cord is attached to the moving antenna structure, so that as the antenna moves in azimuth, the cord raises a weight, which in turn activates the azimuth EMERGENCY LIMIT SWITCH should the antenna travel beyond the  azimuth hard limit switches (see Figure~\ref{fig:AzEmLimSwitch}).

\underline{Elevation}. As the elevation movement is limited to $90^\circ$, the two original NEC elevation ``Emergency Limit Switches" were serviced and restored to their original NEC mounting positions.

\subsubsection{Azimuth direction of travel switches}

Two switches operate a latching relay that provides an indication of the direction of travel as the antenna passes through $0^\circ$ and also inhibits the opposing hard limit switch to allow the antenna to travel beyond $180^\circ$. These switches and the azimuth hard limit switches are all activated by the same striker plate mounted on the moving structure of the antenna, which also solves the problem of the wrap ambiguity.

\subsection{Azimuth cable wrap mechanism}
\label{cable-wrap}

The original cable wrap mechanism only allowed the antenna to travel through $\pm 170^{\circ}$, which is much less than the $\pm 270^{\circ}$ considered to be required for ease of operation of a radio telescope. Therefore a completely new cable wrap mechanism had to be designed and installed and all the original NEC cables were discarded together with the original metal chain.

IGUS, a German company that specialises in plastic energy chains, was approached to provide a suitable design for a replacement of the original NEC chain, which would allow twelve screened control cables, six screened power cables, four coaxial cables and one lightning earth cable - a total of 23 cables of varying size and weight - to be routed through a cable chain system. A design was produced and accepted, and the chain and the chain carrier system was manufactured by IGUS.

The ``Energy Chain" system works by using an inner fixed wall and an outer moving wall; both walls are concentric with each other like two metal cans of different diameters with one inside the other. The plastic chain is looped from one wall to the other and unloops / loops depending on the direction of travel, as shown in Figure~\ref{fig:EnergyChainsystem}.
 
\begin{figure}
    \centering
    \includegraphics[width=7cm]{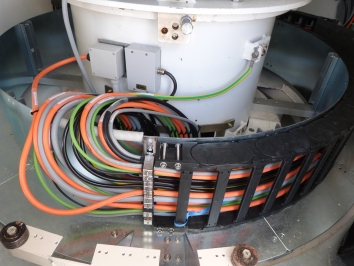}
    \caption{The IGUS Energy chain system in situ. (Images courtesy of Stuart Weston)}
    \label{fig:EnergyChainsystemInsitu}
\end{figure}     

Some difficulty was initially experienced with the chain tending to travel away from the walls when being pushed in the case of the inner wall or being pulled in the case of the outer wall. This was mainly because of the stiffness of the cables going through the tight corner of the loop. A partial solution to the problem was achieved for the case when the chain is wrapping about the inner wall by affixing magnetic strips to the chain vertical struts. These have the effect of magnetically holding the chain against the inner wall while it is being pushed. Unfortunately the magnetic strips were not strong enough to work through the thick coat of paint on the outer wall when the chain was being pulled. However, a solution to this has been achieved by the replacement of the ‘outer guiding plate with a wheeled gantry that ensures no drag on the chain. The new IGUS chain is shown in situ in Figure~\ref{fig:EnergyChainsystemInsitu}.

\subsection{Control system}
\label{sec:ControlSystem}

\begin{table*}
\caption{New parameters after control system replacement.}
\begin{center}
\begin{tabular*}{\textwidth}{@{}ll@{}}
\hline\hline
Description & Specification \\
\hline%
Azimuth Maximum (Tracking and Slewing) Velocity &  0.37 deg/sec or 22.2 deg/min \\
Elevation Maximum (Tracking and Slewing) Velocity & 0.36 deg/sec or 21.6 deg/min \\
Azimuth Acceleration/Deceleration & 0.2 deg/sec/sec \\
Elevation Acceleration/Deceleration & 0.25 deg/sec/sec \\
Azimuth Working Range (as defined by soft limits) & -179.0  to 354.0 deg \\
Elevation Working Range (as defined by soft limits) & 6.0 to 90.1 deg \\
\hline\hline
\end{tabular*}
\end{center}
\label{tab:cont_sys_params}
\end{table*}

The original antenna control system was only designed to meet the requirements of the Earth Station's telecommunications, which primarily meant pointing at geostationary telecommunication satellites with very limited motion/tracking.  In order to function as an efficient radio telescope, the new control system should be able to provide a much greater variety of drive modes. Examples of these are scanning the telescope relative to the position of a source whilst continuing to track it in one of several different astronomical co-ordinate systems; quickly changing from one object to another; finding the optimum path from one radio source to another etc. A modern interface to some form of control panel, so that the telescope can be driven either manually or under the control of a general purpose modern-day computer, programmed to provide a complicated observing schedule, is a basic requirement, which has been mostly implemented. Parameters of the new drive capability, to be compared with the originals given in Table \ref{tab:nec_spec}, are given in Table \ref{tab:cont_sys_params}.

Originally the antenna was fitted with two sets of motors; large (11~kW) induction motors for slewing and small DC servomotors with extra gearing for tracking the small daily motions of geostationary satellites.  Two pairs of motors of each type were used on each axis to provide anti-backlash torque, and a system of clutches selected between the slew and track motors.  

For tracking over the full speed range of the antenna the large induction motors were replaced with brushless 55~Nm (nominal torque) AC servomotors with optical shaft encoders.  (The old DC servomotors are still present but have been permanently disengaged from the drive chain.)  There are no longer separate modes for slewing and tracking, and the antenna is always under closed loop position control.   

A new Integrated Antenna Controller (IAC) was developed and installed by M P Godwin Ltd., UK. It provides all the functionality of the motor inverter drives and a high-level antenna control unit, and resides in a single cabinet located in the equipment room beneath the antenna.  The IAC uses only commercial, off-the-shelf, process-control hardware and follows a design that has been successfully used on antennas ranging in size from one metre to tens of metres in diameter for low Earth-orbit satellites, geostationary satellites and astronomical tracking applications.  

The IAC has an inverter drive for each of the four servomotors and inbuilt drive firmware handles the motor current and speed control loops.  However, the drives provide extensive capability for user programmability, and this allows the IAC to run custom software for different antenna control applications.  The position control loop algorithms are implemented in these areas, as well as motion controller algorithms specially designed for careful control of the accelerating forces and jerk applied to the antenna structure.  In this configuration, all elements of the position control algorithm are synchronous and timing jitter is therefore not a problem.  Synchronous communication between drives is another important feature for sending the control demands to the pairs of motors that drive each axis to limit mechanical backlash. 

A range of functions specific to the use of the antenna as a radio telescope have been incorporated into the IAC's application software including:

\begin{itemize}
  \item A clock (set and regulated from network time)
  \item Telescope Pointing Error Correction using a standard nine term error model \citep{TDA_Report_42_88}
  \item Correction for atmospheric refraction
  \item Accepts position commands in both the Horizontal (Az, El) and Equatorial (RA, Dec) coordinate system
  \item Interpolator for tracking from time tagged data (2 x 2000 point arrays)
  \item Command Scheduler 
  \item Monitoring and diagnostics
\end{itemize}

The application software and the position control algorithms cycle every 4\,ms. Remote communication with the IAC is via a 10/100 Mbps Ethernet interface and optical fibre.  Control and monitoring uses the well-proven Modbus TCP/IP (Industrial Ethernet) master slave protocol and multiple clients are supported.  The antenna can also be driven locally from the IAC for maintenance purposes.

One of the challenges in designing a new control system for this antenna was very limited knowledge of its mechanical characteristics such as stiffness, inertias, drive train efficiencies, wind loads, etc.  However commissioning tests have shown the system to be stable with a servo accuracy of better than one millidegree under light wind conditions.   Further operational experience is awaited regarding wind gust performance, but sufficient margin is available to tighten the system response further if required. 

\subsection{Cleaning and maintenance }
\label{cleaning}

\begin{figure}[htbp]
\centering
\begin{subfigure}
  \centering
  \includegraphics[width=0.15\textwidth]{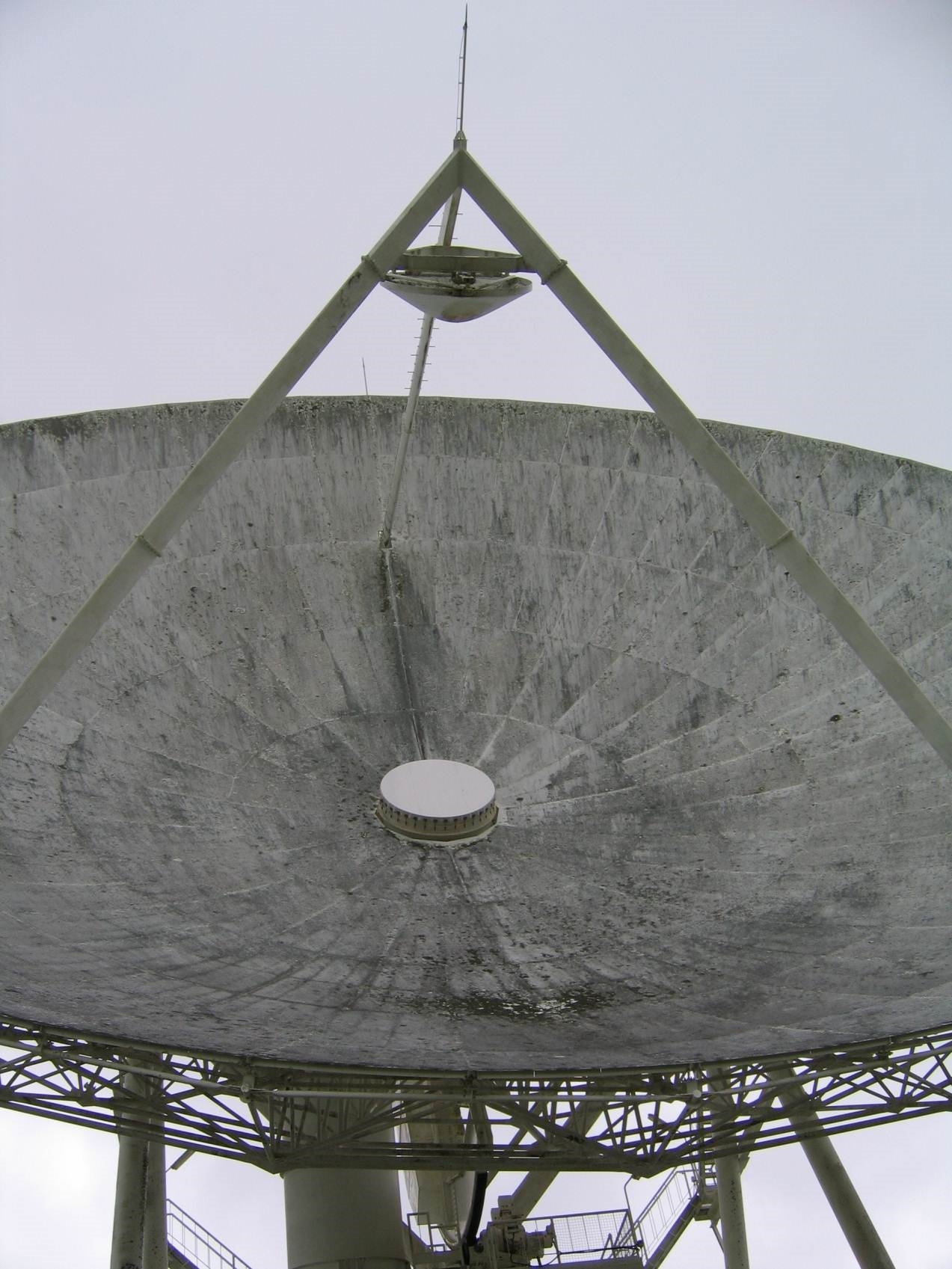}
   \label{fig:sub1}
\end{subfigure}%
\begin{subfigure}
  \centering
  \includegraphics[width=0.265\textwidth]{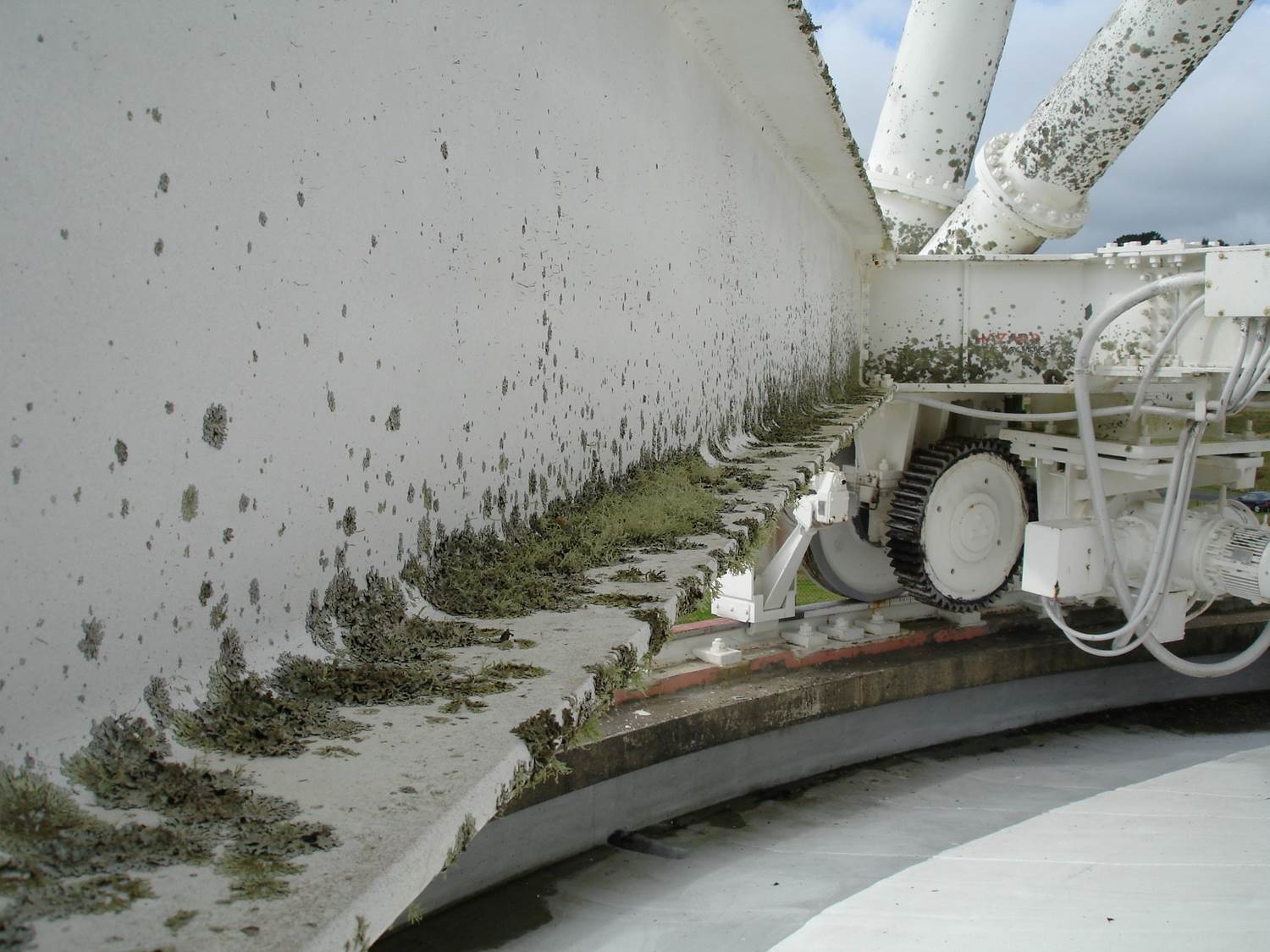}
  \label{fig:sub2}
\end{subfigure}
\\
\begin{subfigure}
  \centering
  \includegraphics[width=0.15\textwidth]{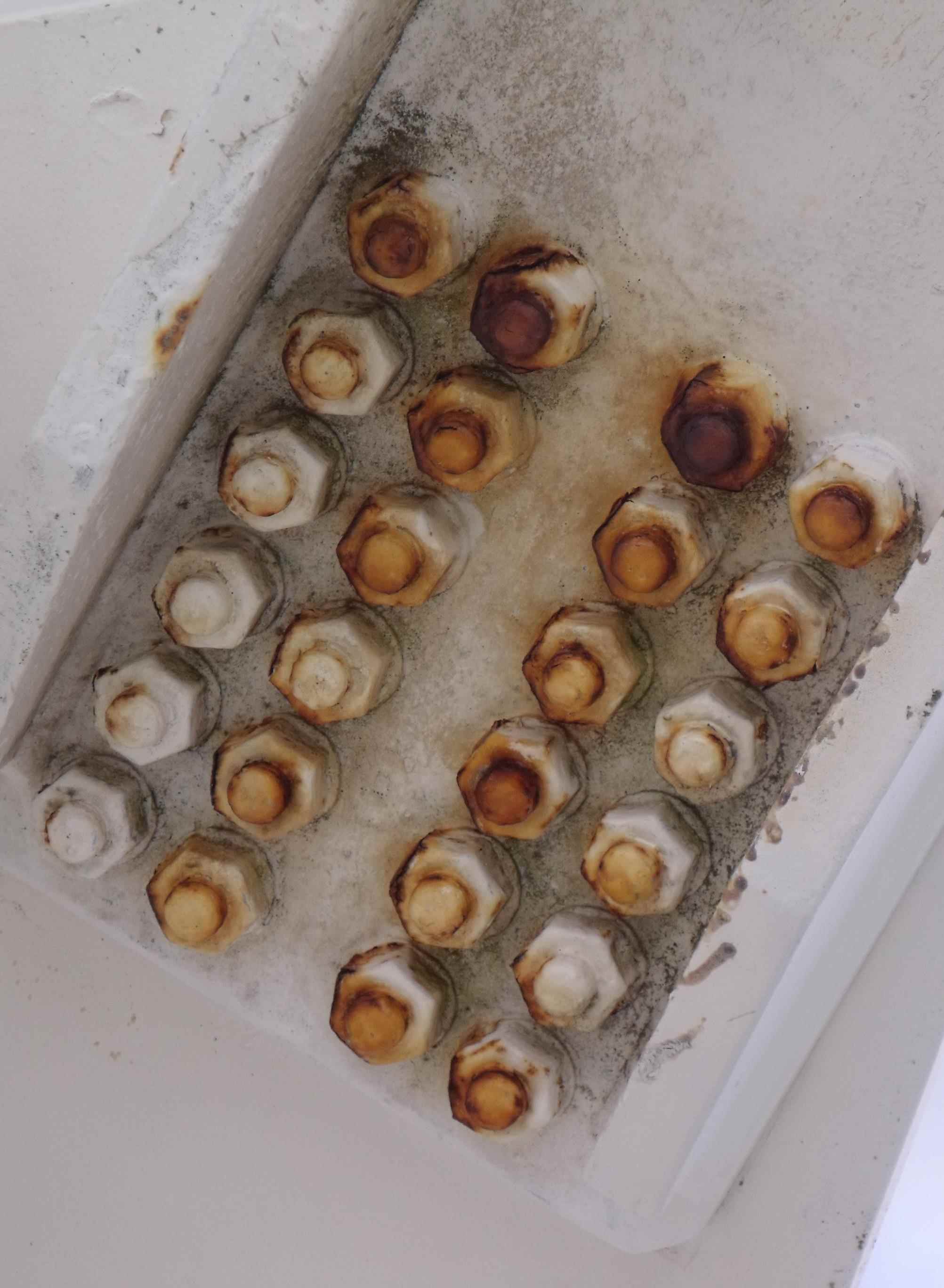}
   \label{fig:sub3}
\end{subfigure}%
\begin{subfigure}
  \centering
  \includegraphics[width=0.265\textwidth]{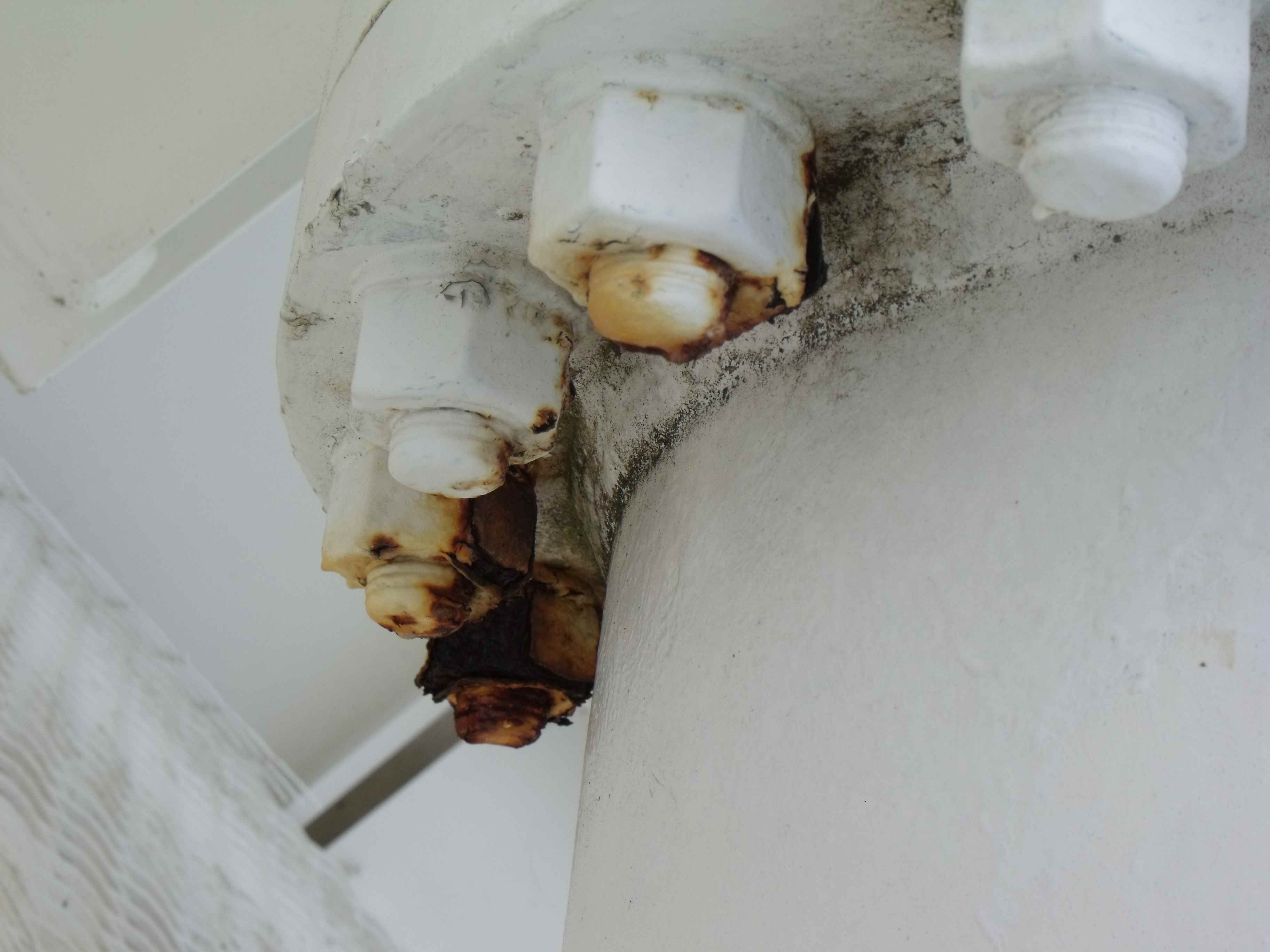}
  \label{fig:sub4}
\end{subfigure}
\caption{Photos of the 30-m antenna (before cleaning).(Images courtesy of Stuart Weston)}
\label{fig:30mDirty}
\end{figure}

The telescope structure is only 5~km from the sea to the east. Thus, salt and corrosion are an issue and Telecom NZ had ceased maintenance some time before passing the dish to AUT as they anticipated demolition. A backlog of several years of maintenance and cleaning (e.g. replacement of rusty bolts, surface cleaning and painting etc.) had therefore built up. Figure~\ref{fig:30mDirty} shows the initial state of the main reflector surface, some bolts and part of the main support structure.  A local rigging contractor has been engaged to assist in a programme of bolt replacement and general maintenance/repairs of the dish structure, which was introduced in 2012 and continues. At the time of writing the backlog has been reduced to zero, but the general maintenance and on-going repairs continue following a schedule. 

\section{SURVEY OF THE MAIN REFLECTOR SURFACES}
\label{surface}

\begin{figure*}
\centering
\begin{tabular}{c}
\includegraphics[width=15cm]{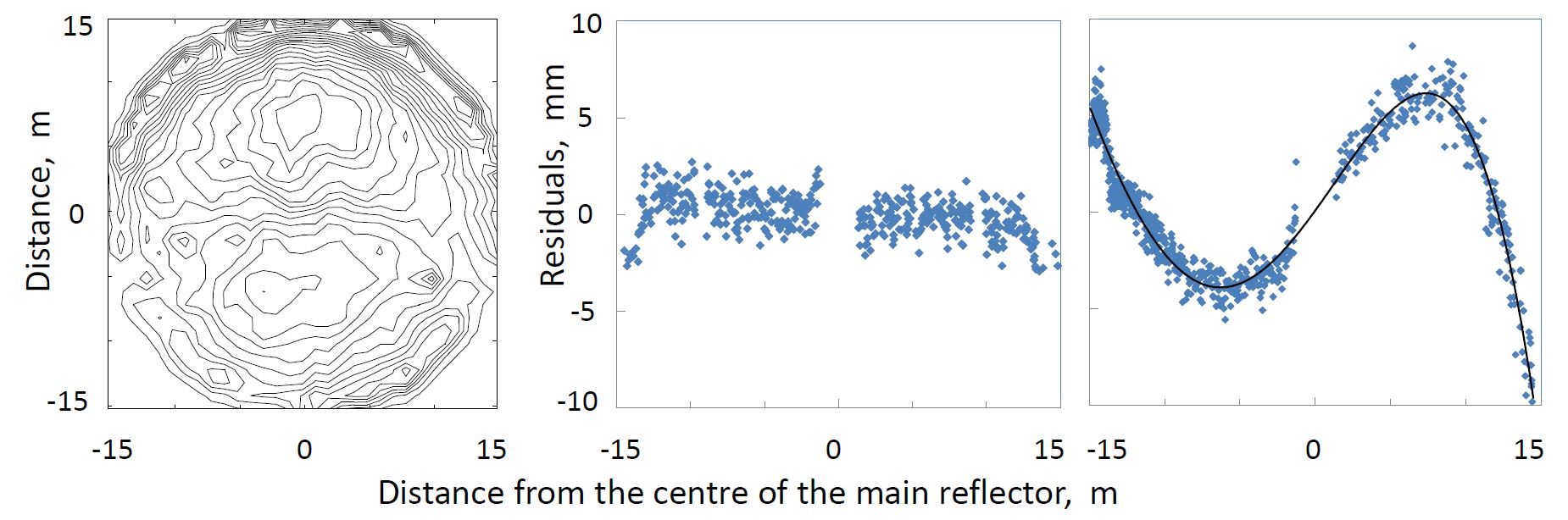}
\end{tabular}
\caption{Measurements of the main reflector surface at the lowest elevation angle of $6\,^{\circ}$: (left) the contours for residuals; (centre) the residuals for a cross-section taken through the main reflector centre along the horizontal direction; (right) same along the vertical direction.}
\label{fig:dish_surface}
\end{figure*}

The accuracy of the main reflector surface has been investigated using a FARO Laser Scanner provided by Synergy Positioning Systems Ltd. The initial scanning was conducted from the ground with the dish positioned at an elevation of $6\,^{\circ}$ and measurement points defined on the reflector surface with a 1\,mm separation.  The rms of the distance measurement was 1\,mm \citep{fig2013_ts08c}. The data were processed and a comparison of the measured surface with a theoretical surface has been made. The results of the data processing revealed a noticeable gravitational deformation of the antenna in the vertical direction with the telescope pointing at this elevation of $6\,^{\circ}$.

Figure~\ref{fig:dish_surface} (left) shows a map of the deviations of the main reflector surface from the theoretical surface at its lowest elevation angle of $6\,^{\circ}$. Figures~\ref{fig:dish_surface} (centre) and (right) show plots of the residuals for horizontal and vertical cross-sections through the reflector's centre respectively. The rms deviation of the whole surface from the best theoretical surface is $\sim 3.5$\,mm. For the horizontal cross-section that is not affected by the gravitational deformation, the rms is $\sim 1.0$\,mm, which is in agreement with the accuracy of the laser scanner (1 mm). However, the rms deviation along the vertical cross-section through the main reflector centre is $\sim 5$\,mm.  A more detailed investigation of the gravitational deformation and its dependence on the antenna elevation is in preparation.

\section{INSTRUMENTS}

\subsection{Receiver}

The radio telescope is equipped with a beam waveguide system bringing the RF signal down into a feedhorn mounted vertically within the building underneath the telescope. As originally installed in 1982, the feed was configured for C-band satellite communications with a receive signal downlink in the range $3.7\,-\,4.2$\,GHz and a transmit uplink in the range $5.925\,-\,6.425$\,GHz. The orthomode transducer (OMT), diplexer and all other components associated with this telecommunications use have been removed, leaving just a small section of tapered waveguide attached to the mode converter (output) section of the feed.  

\begin{figure}[!h]
   \begin{center}
     \includegraphics[scale=0.1]{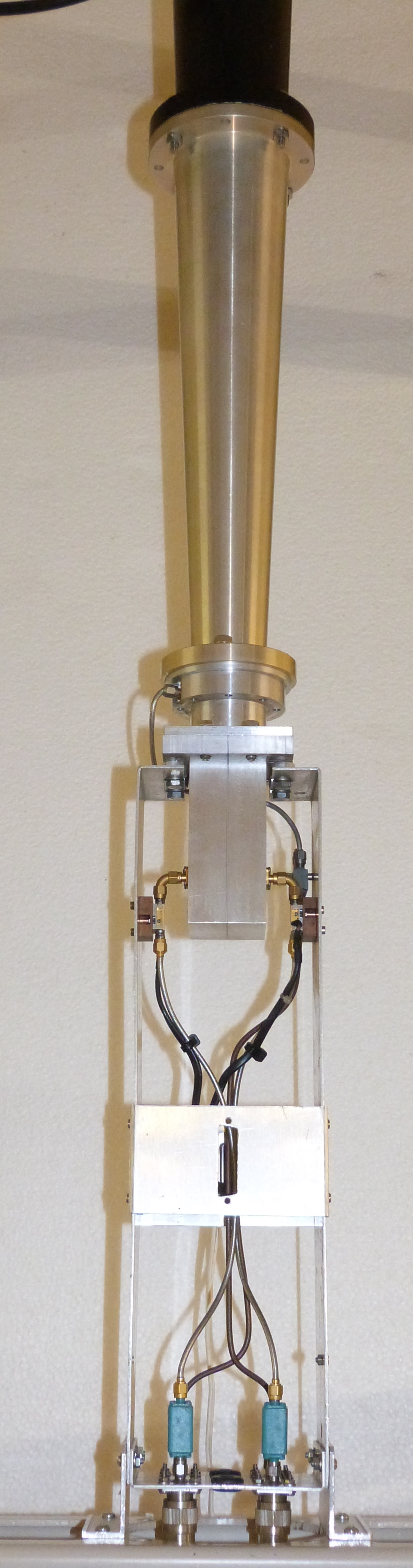}
   \end{center}
   \caption{The C-band receiver with the new feedhorn transition unit manufactured by BAE Systems Australia. (Image courtesy of Stuart Weston)}
   \label{fig:jb-cband-receiver}
\end{figure}

A waveguide transition, designed and manufactured by BAE Systems Australia, has been installed  to provide the connection between the feed (i.e. the remaining small section of tapered waveguide) and the OMT input of an uncooled 6~GHz receiver donated by Jodrell Bank Observatory (Figure~\ref{fig:jb-cband-receiver}). 

The receiver, with a centre frequency of $\sim 6.6$\,GHz and a bandwidth of $\sim 300$\,MHz, uses HEMT amplifiers to provide low noise gain, and a stepped orthomode transducer to provide dual circular polarisation. A downconverter, also provided by JBO, is used to mix the RF signals with a nominal $\sim 5.8$\,GHz local oscillator, locked to the hydrogen maser, to produce IF output frequencies in the range $750 - 1050$\,MHz. After further amplification and filtering, these IF signals are then input to the DBBC analogue to digital modules described in Section \ref{sec:dbbc}.  The receiver also has provision for injection of both amplitude and phase calibration signals, the latter specifically for use during VLBI sessions. A very preliminary calibration of the system's performance has been undertaken by recording the receiver output power when pointing ON and OFF the source Taurus A.  However, because of the weather, the results have not yet been sufficiently consistent to report and proper system temperature measurements still need to be made using `hot and `cold' (liquid nitrogen) loads and standard calibration sources in good weather conditions.

\subsection{Field System}

To maintain consistency and to reduce the support effort required, it has been decided to use the same control and scheduling system software for both the 12-m and 30-m antennas, i.e. the NASA / Haystack Field System \citep{fieldsystem08072014}. However, there are a few necessary differences in such modules as ``The Antenna Control interface programme'' ( ``$antcn$'') because of the physical differences between the two antennas, e.g. the number of motors, etc. Although the antenna controller itself has a built-in 9 point coefficient pointing model as described in section \ref{sec:ControlSystem}, it has been decided to use the Field System pointing routines and pointing model \citep{FSPointingModel} because of the automation provided within the Field System for its support and maintenance, and also its well-proven use with the NASA deep space network.


\begin{figure}[!h]
\centering
\begin{tabular}{c}
\includegraphics[width=8.5cm]{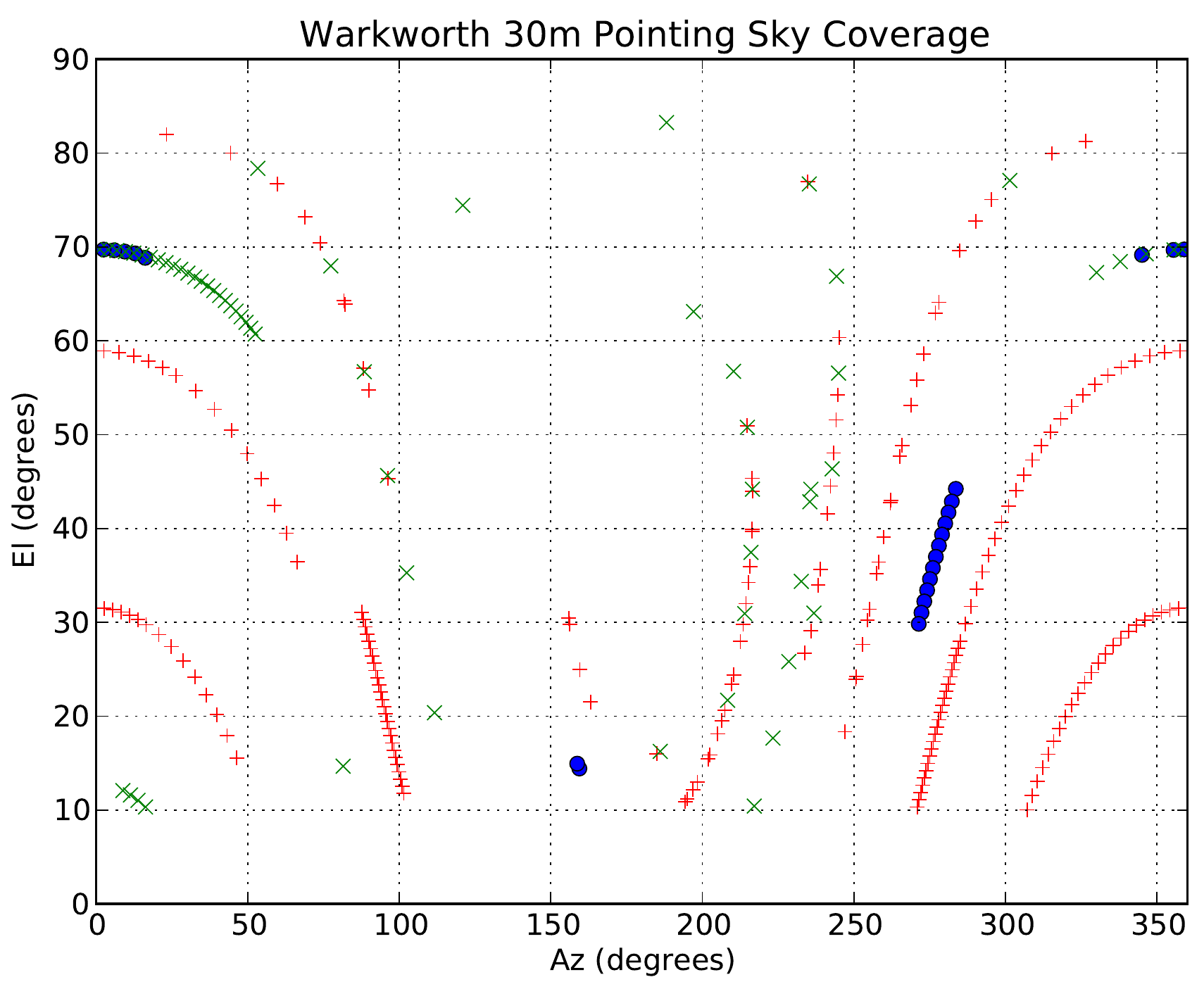}
\end{tabular}
\caption{First pointing sky coverage, the symbols are the same from Figure~\ref{fig:pointing-errors}. The  blue filled circles are before any model; red $+$ for first model; green x current model.}
\label{fig:pointing-sky-coverage}
\end{figure}

\begin{figure}[!h]
\centering
\begin{tabular}{c}
\includegraphics[width=8.5cm]{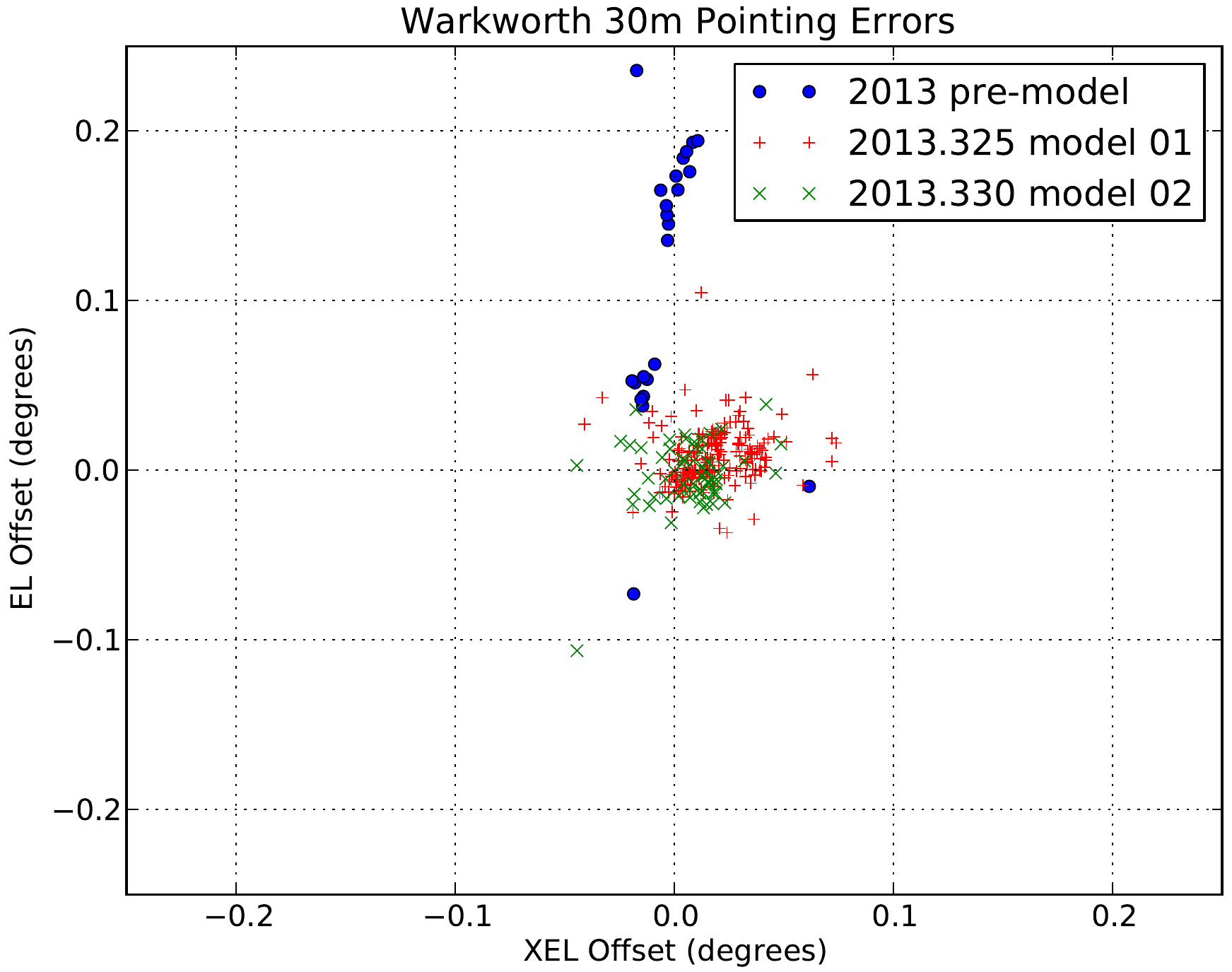}
\end{tabular}
\caption{Comparison of the first pointing, over several days in 2013 building new models using the Field System "acquire" and "fivpt". This is a plot of the EL and XEL offsets in degrees: blue filled circles are before any model; red $+$ for first model; green x current model.}
\label{fig:pointing-errors}
\end{figure}


It can sometimes take up to 15 min to move between sources requested by the pointing algorithm (``acquire") of the Field System because of the low angular velocity (22 deg/min) of the 30-m antenna. To minimise the time spent slewing (and consequently to maximise the time on pointing sources), a strategy whereby the source catalogue was split into two was adopted. Whilst arguably not optimal, it was found that the simple strategy of dividing the catalogue into subsets of northern (Dec~$\geq 0^{\circ}$) and southern (Dec~$< 0^{\circ}$) sources, and scanning the northern subset and then the southern subset, provided a worthwhile reduction in the time to acquire sufficient data to generate a statistically robust pointing model.  

Initial pointing measurements using an Agilent U2000A RF power meter and only northern sources were undertaken to determine approximate values for the basic offsets and coefficients. Observations using both northern and southern pointing catalogues were then made for a 6hr period to give the sky coverage shown in Figure~\ref{fig:pointing-sky-coverage} (red plus signs). After corrections to the pointing model based on the results of the above measurements had been made, a third run was undertaken to verify the offsets and coefficients produced from the second pointing sequence.  The results following each pointing run are shown in Figure~\ref{fig:pointing-errors}, where incremental improvements in the pointing model can be seen.  The final model has rms values in azimuth (cross-elevation) of $1.1\,^{\prime}$ and in elevation of $1.2\,^{\prime}$

Eventually the DBBC (see Section \ref{sec:dbbc}) will be used for automated pointing with the FS module ``$acquire$'', as now used on the 12-m. Regular monitoring of the pointing model will also need to be carried out to see if there is any drift in either the azimuth or elevation offsets.

\subsection{Digital Base Band Converter}
\label{sec:dbbc}

The 30-m radio telescope has been equipped with a Digital Base Band Converter (DBBC) supplied by HatLab \citep{GinoIVS2010}. Together with a MK5B recorder (see section \ref{sec:recorder}), one of these DBBCs has been in use with the Warkworth 12-m telescope since 2009. Originally developed as part of a RadioNet project started in 2005 to provide a digital replacement for the analogue converters in the VLBI acquisition systems of the EVN radio observatories \citep{VLBI2010}, the DBBCs are now in widespread international use. 

The analogue input section of a DBBC is fitted with four RF modules (IFs), each with four separate RF inputs and a programmable switch that allows a user to specify which of the four inputs is connected to the digitisation and processing sections of the DBBC. Each IF module is equipped with a set of 4 filters (0.01-512, 512-1024, 1024-1536, 1536-2048 MHz) that can be set by software to define the 512~MHz analogue input range the module will accept. At the time of writing, the DBBC is configured with two IF modules, each connected to a single polarisation of the dual polarisation receiver (RCP and LCP).  It is envisaged that all four IF modules will be required when the antenna is equipped with either a dual frequency or a broadband receiver.

Now that support for the DBBC has been incorporated into the NASA Field System, the DBBC will be configured through the use of procedure files generated when VLBI schedule files are processed on the station Field System.  

\subsection{Recorder}
\label{sec:recorder}
The recording system is a Mark5B \citep{whitney_2006} connected to the DBBC via a VLBI Standard Interface (VSI). These units, purpose built for VLBI, can also be used for recording single dish experiments. A separate stand-alone Mk5C has also been obtained for e-transfers of the data to the end user or correlator, thus not tying up the recording unit with data transfers and therefore preventing further observations.  The FiLA10G interface of the DBBC will eventually be used to stream data to the Mk5C recorder once further work to establish a station-wide 10G Ethernet LAN has been completed.

\section{WARKWORTH OBSERVATORY}

\subsection{Site Survey and tie to the GNSS station}

\begin{figure}[H]
    \centering
    \includegraphics[width=8cm]{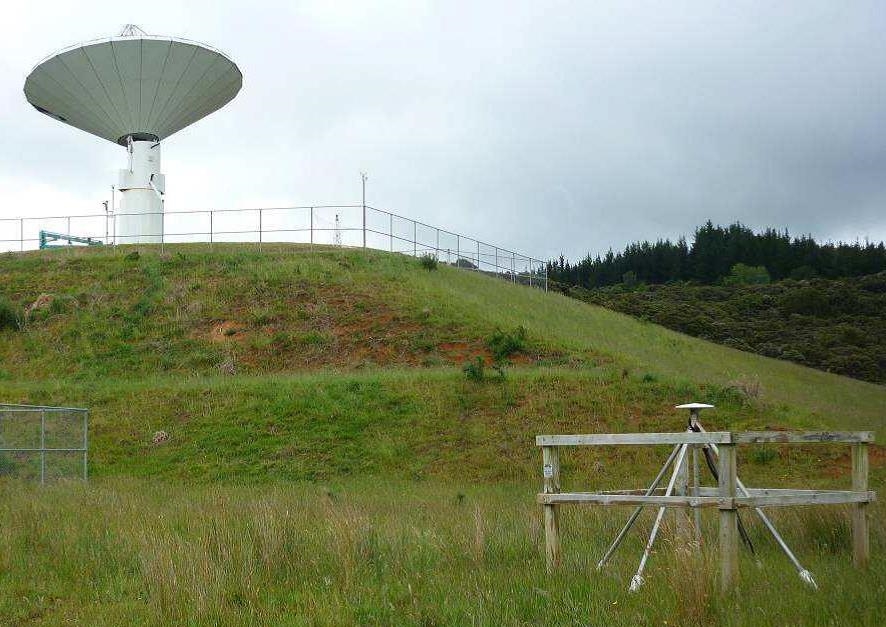}
    \caption{Collocated space geodesy facilities:  the GNSS base station at Warkworth (WARK) and the IVS network station WARK12M (12m radio telescope). (Image courtesy of Sergei Gulyaev)}
    \label{fig:GNSSBaseStation}
\end{figure}   

Important parts of the 30-m antenna science case are space geodesy and participation in geodetic projects, such as AuScope \citep{AuScope} and IVS \citep{IVS}.  A Global Navigation Satellite System (GNSS) base station, WARK, is part of the International GNSS Service (IGS) global network and is co-located with the 12-m geodetic antenna  as shown in Figure~\ref{fig:GNSSBaseStation}. A tie survey of the 12-m and GNSS station has been conducted by Land Information NZ (LINZ) at the end of 2012 \citep{gentle_2013} and a plan for a geodetic survey of the 30-m antenna is being discussed with LINZ. 

The WRAO has a Symmetricom Hydrogen maser as the primary frequency standard \citep{IVS_Annual_Report_2008}, which is housed in a temperature controlled environment within the 12m antenna control cabin. The time, derived from GPS, and the frequency standard are transferred from the 12-m control cabin to the 30-m via a fibre link between the two antennas using a Symmetricom fibre-optic distribution amplifier and a fibre-optic receiver to generate the standard signals for use by equipment in the 30-m building.

\subsection{Networking}

It is envisaged that the 30-m radio telescope will regularly participate in VLBI and eVLBI observations, which means that the issue of data transfer is a high priority. In 2010 the Research and Education Advanced Network NZ (REANNZ) installed a 1~Gbps point of presence (Giga-PoP) at the WRAO to provide high-speed, data transfer from the 12-m telescope. This network was upgraded on 28th April 2014 from 1 Gbps to 10 Gbps, and shortly after this the REANNZ international links were upgraded to 40Gbps. This should enable e-VLBI operations with 1 Gbps speeds or greater (i.e. 16 channels each with 16 MHz bandwidth and 2-bit resolution).

\section{SCIENCE WITH THE NZ 30-m RADIO TELESCOPE}
\label{science}
Assuming frequencies of operation from 1.3~GHz to 10~GHz (perhaps higher) would be possible with the 30-m telescope, a variety of astronomical and geodetic research could be undertaken, both using the telescope as a stand-alone single dish and also in conjunction with other telescopes as a contributor to the interferometer networks. Dependent upon the frequencies of operation actually available, it could contribute significantly to the IVS as well as the geodetic AuScope network. However, it is to be noted that the geodetic IVS/AuScope observations would require simultaneous S- and X-band measurements. 

\subsection{VLBI Observations}

Astrophysically, the 30-m radio telescope can contribute to both continuum and spectral line VLBI observations in the study of both galactic and extragalactic radio sources. The telescope will be a valuable addition to the Australian Long Baseline Array (LBA), and to the Asia-Pacific (APT) and European (EVN) VLBI networks. 

Under the heading of extragalactic sources likely to dominate continuum observations come active galaxies. In this category are active galactic nuclei (AGN), both radio-quiet (such as Seyfert galaxies, low-ionization nuclear emission-line regions (LINERs) and radio-quiet quasars) and radio-loud (blazars, radio-loud quasars and radio galaxies), especially at high redshift.   Also in this category are starburst galaxies, luminous and ultra-luminous infrared galaxies (LIRGs and ULIRGs), and gravitational lenses. The formation of supermassive black holes and their development in AGN, particularly at high redshift (i.e. in the early Universe) is currently of particular interest.  In comparatively nearby starburst galaxies, LIRGs and ULIRGs, the occurrence and development / expansions of supernovae and their remnants, especially in objects such as ARP\,220 and IC\,883, which are now being shown to contain AGN, are attracting considerable research. Ultra-steep spectrum (USS) radio sources also seem to be near the top of lists in searches for high-redshift radio galaxies (HzRG), and images of their structures (core, jets  and lobes) are of importance when considering the formation of black holes.
 
For galactic sources, areas of special interest include the light curves and images of the expansions of recurrent novae, micro-quasars and transients.

Perhaps perceived at present to be the most important VLBI spectral line observations will be those of external galaxies at intermediate and low redshifts in the 21-cm hydrogen line, and of the OH maser lines at 1.6~GHz and the methanol maser line at 6.7~GHz in star-forming regions in our own galaxy.  VLBI OH polarisation observations in the 4 ground state OH lines at 1612, 1665, 1667 and 1720~MHz should allow the direct measurement of magnetic fields in the star-forming regions using the Zeeman effect. Follow-up VLBI observations of the methanol maser sources at 6.7~GHz discovered in the Parkes multi-beam survey should provide further information on expanding methanol shells and  variability of these maser sources with time. The study of OH megamasers in external galaxies and of small-scale structure in our own galaxy from 21-cm absorption of extra galactic sources will undoubtedly continue. 

 
\subsection{Single-Dish Observations}

In addition to the contribution that the 30-m telescope can make to VLBI observations, it is a valuable instrument in its own right for single-dish measurements, providing a six fold increase in collecting area compared with that of the existing 12-m Warkworth antenna.

At L-band frequencies pulsar timing and polarisation measurements can lead to a better understanding of glitches in pulse periods (and therefore of the neutron stars themselves), of inter-pulses and the radiation mechanism and the variability of the pulses themselves.  Total power observations of interplanetary scintillation (IPS), conducted simultaneously and in conjunction with other individual telescopes, can provide considerable information on the solar wind and the interplanetary medium itself.

At C-band frequencies, light curves of recurrent novae, symbiotic variables, micro-quasars and transients can be observed from their early stages. 

Assuming almost unlimited integration time, important studies of radio recombination lines of hydrogen, helium and carbon can be conducted at all available wavelengths in the various HII regions (star-forming, planetary nebulae and ultra-compact) to determine their parameters such as their densities, temperatures, dynamics and abundances. This is particularly the case for objects not accessible to the northern hemisphere radio telescopes.     

Maser spectral line observations in a single-dish mode will include 1.6~GHz OH lines in outbursts and monitoring of light curves of OH masers from semi-regular and other variable stars. At C-band frequencies, spectral line observations of the OH maser lines at 6030 and 6035~MHz and methanol masers at 6.7~GHz are important for the identification and study of star-forming regions.  The existence of methanol masers often indicates the presence of a star-forming region, when it might otherwise not be visible. The 30-m radio telescope in New Zealand can be effectively used for the monitoring of variability of methanol maser sources found by the Parkes multi-beam survey.


\section{CONCLUSION}

\begin{figure}[!h]
\centering
\begin{tabular}{c}
\includegraphics[width=8cm]{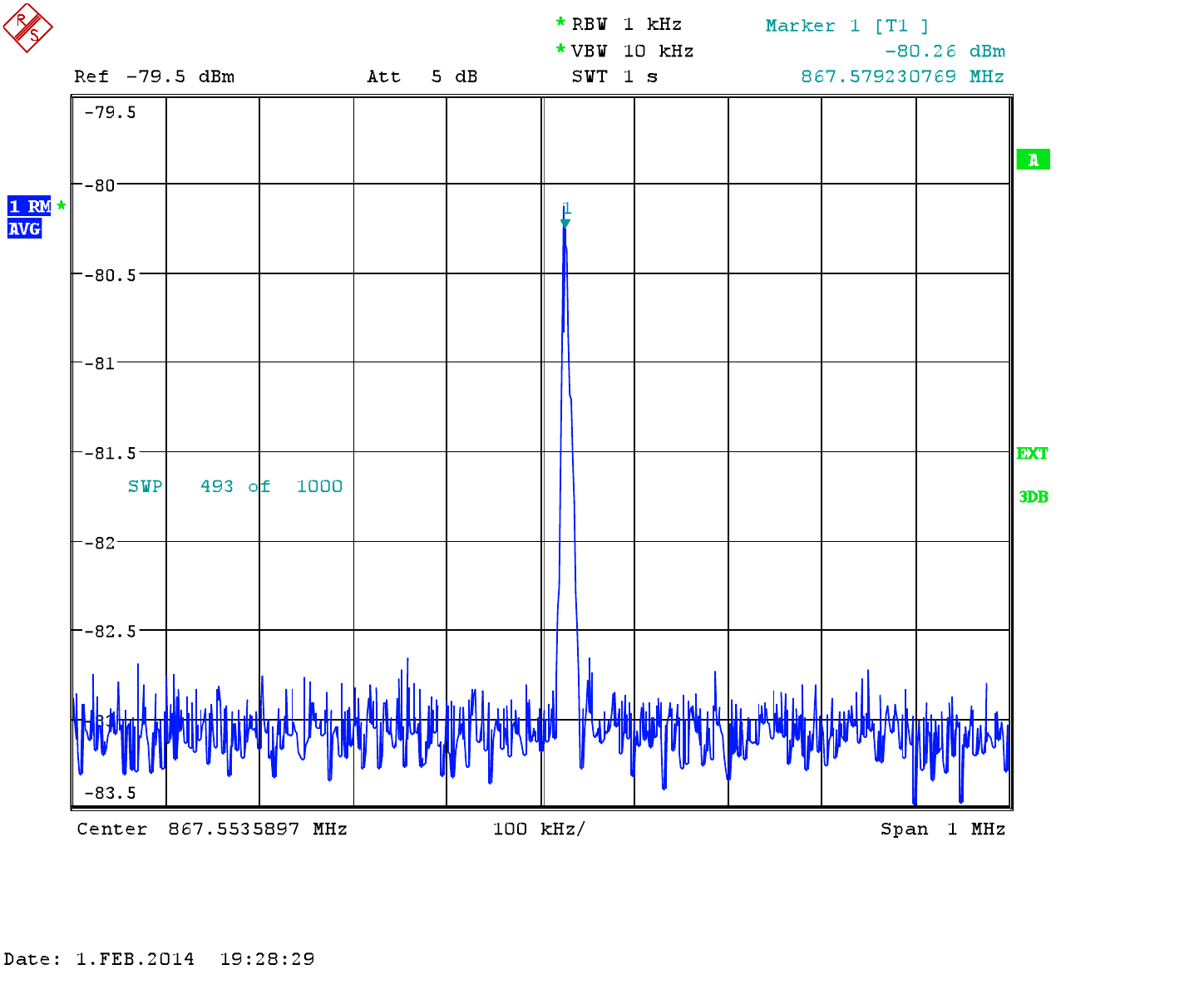}
\end{tabular}
\caption{The "First Light": the spectrum of the galactic Methanol Maser source G188.95+0.89 near 6.7~GHz.}
\label{fig:maser_g188}
\end{figure}

The  ``official" First Light of the Warkworth 30-m radio telescope took place on the 4th of July 2014. Figure~\ref{fig:maser_g188} shows the spectrum of a galactic methanol maser source obtained at that time. On the 11th of December 2014, the first geodetic and astrophysical scientific VLBI experiment was conducted; the results will be presented in (Petrov et al., 2015, in preparation). 

This 30-m antenna, with a large surface area, adds significantly to New Zealand`s capability in radio astronomy and is a highly sensitive instrument capable of significant single dish work. In addition, the inclusion of the Warkworth 30-m antenna will greatly enhance the LBA, especially as it contributes to the longest baselines. This is also true when used in conjunction with other Asia-Pacific antennas.

A significant amount of work has been undertaken during the last few years to convert the telecommunications antenna into a radio telescope capable of conducting important research in astronomy and astrophysics. Significant changes had to be made to the mechanical, electrical and electronic systems of the facility to enable professional radio astronomical operations both in single-dish and VLBI modes. 

Further investigation of the gravitational deformation of the main reflector surface is required in order to clarify the ultimate capability of the telescope at centimetre wavelengths. 

A next step in the improvement of the radio telescope sensitivity and enhancement of its capabilities and usage will be the installation of a cryogenically cooled C-band receiver. Solutions for S- and X-band and possibly lower (L-band) and higher frequency (K-band) capability will be sought. The S- and X-band receivers, if implemented, should enable use of the 30-m radio telescope for geodetic research as a contributor to the IVS and AuScope, and in cooperation with Geoscience Australia and Land Information New Zealand.  

The Warkworth Radio Astronomical Observatory is operated by the AUT University where the Astronomy Major was introduced in the School of Computer and Mathematical Sciences in 2010. The Observatory plays an important role in the preparation of both undergraduate and post graduate students, providing a facility for laboratory work and students' research projects.
  
The staff and postgraduate students of the Institute for Radio Astronomy and Space Research are involved in the design work of the Square Kilometre Array, mainly in its Central Signal Processor and Science Data Processor work packages. Potentially, the observatory and its facilities, including its supercomputing centre can be used as a testbed for some of the SKA design work conducted in New Zealand.

\begin{acknowledgements}
We would like to thank the many parties that have helped to make this conversion possible. This is not an exhaustive list but includes NZ Telecom; BAE Systems, Australia; the University of Manchester, Jodrell Bank Observatory; the Australia Telescope National Facility, which is funded by the Commonwealth of Australia for operation as a National Facility managed by CSIRO; and perhaps above all the Vice-Chancellor and most Senior Academic Staff of Auckland University of Technology, who have invested in the project.

Stuart Weston is a recipient of a CSIRO Astronomy \& Space Science, Australia Telescope National Facility Graduate Research Scholarship.

We would like to thank the reviewer for their insightful comments on the paper, as these comments have
led us to an improvement of the work.
\end{acknowledgements}


\bibliographystyle{apj}
\bibliography{mnemonic,paperbib}


\clearpage

\end{document}